# Tuning the Kondo effect with back and side gates – Application to carbon nanotube superconducting quantum interference devices and pi-junctions

**J.-P. Cleuziou[1], W. Wernsdorfer[2*], V. Bouchiat[3], Th. Ondarçuhu[1], M. Monthioux[1]**

[1]*Centre d'Elaboration des Matériaux et d'Etudes Structurales, CEMES-CNRS, 29 rue Jeanne Marvig, 31055 Toulouse Cedex 4, France*
[2]*Laboratoire L. Néel, LLN-CNRS, associé à l'UJF, BP 166, 38042 Grenoble Cedex 9, France*
[3]*Centre de Recherches sur les Très Basses Températures, CRTBT-CNRS, associé à l'UJF, BP 166, 38042 Grenoble Cedex 9, France*

**We recently presented the first superconducting quantum interference device (SQUID) with single-walled carbon nanotube (CNT) Josephson junctions [1: J. P. Cleuziou, W. Wernsdorfer, V. Bouchiat, T. Ondarçuhu and M. Monthioux, Nature Nanotech. 1, 53, (2006).], http://www.nature.com/nnano/journal/v1/n1/pdf/nnano.2006.54.pdf . We showed that quantum confinement in each junction induces a discrete quantum dot (QD) energy level structure, which can be controlled with two lateral electrostatic gates. In addition, a backgate electrode can vary the transparency of the QD barriers, thus permitting to change the hybridization of the QD states with the superconducting contacts. This technique is further illustrated in this additional supporting material where we show that the Kondo coupling for a given resonance can be continuously tuned by varying the backgate voltage. It allowed us to show [1] that CNT Josephson junctions can be used as gate-controlled pi-junctions, that is, the sign of the current-phase relation across the CNT junctions can be tuned with a gate voltage.**

The problem of a quantum dot coupled to superconducting electrodes has been the subject of intense studies in the last decade. Among the expected new effects, it was predicted that a reverse superconducting current (Josephson current) would take place in a junction involving tunnelling through a quantum dot (QD) populated with an odd number of electrons[2-7]. Such an effect, in which the minimum energy state of one of the Josephson junction is obtained for a phase difference of π instead of 0, has been the object of intense studies in the last decade. It has been reported experimentally as a consequence of d-wave superconductivity[8] when tunnelling occurs through a ferromagnetic layer[9,10]. Local control of the sign of a Josephson current was also proven possible[11] by tuning the local density of state of a SNS junction[12]. For a strong Kondo effect of a QD populated with an odd number of electrons, the Josephson coupling is expected to be positive (0-junction) since the localized spin is screened due to the Kondo effect. On the other hand, for a weak Kondo effect, the large on-site interaction only allows the electrons in a Cooper pair to tunnel one by one via virtual processes in which the spin ordering of the Cooper pair is reversed, that is, the Cooper pair wave function is multiplied by a phase factor $e^{i\pi}$, leading to a negative Josephson coupling (π-junction).

In order to demonstrate that a carbon nanotube (CNT) Josephson junction can be used as gate-controlled pi-junction, we propose to use backgate and sidegate electrodes, which can vary the transparency of the QD barriers, thus permitting to change the hybridization of the QD states with the superconducting contacts. It permits therefore to vary the Kondo effect, thus permitting to find the right Kondo interaction, which allows us to use the CNT Josephson junction as a gate-controlled pi-junction.

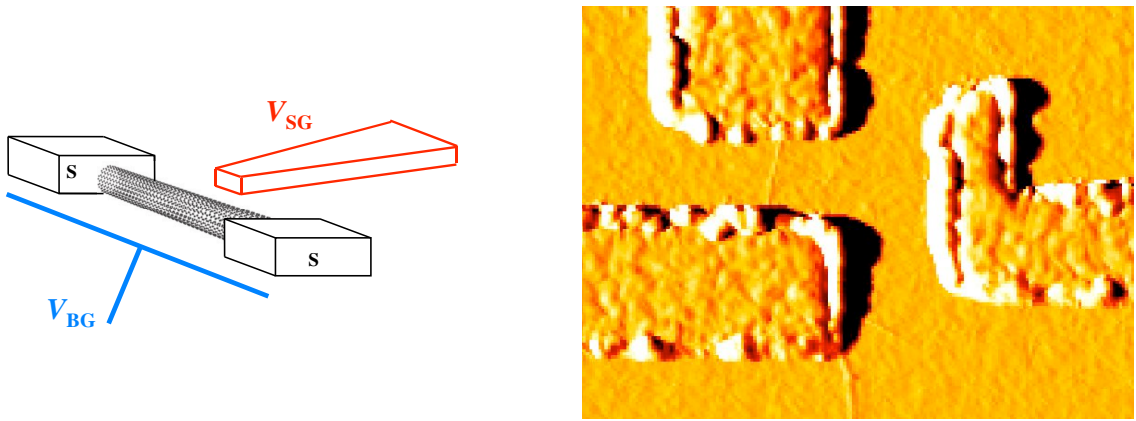

**Figure 1. a** Schematics of the CNT-junction with a nanotube junction, which can be tuned with the gate voltages $V_{BG}$ and $V_{SG}$ corresponding respectively to the back- and sidegate electrodes. **b** Typical device geometry of the CNT-junction with a lateral gate G. The atomic force microscope (AFM) image shows the junction, which is interrupted by a carbon nanotube (CNT) Josephson junctions with a length of about 200 nm. The single-walled carbon nanotube with a diameter of about 1 nm was located using AFM and Pd/Al (3/50 nm) aligned electrodes are deposited over the tube using electron-beam lithography.

In order to build a CNT Josephson junction as presented in Fig. 1b, we started from a degenerately n-doped silicon substrate with 350 nm thick thermally grown $SiO_2$ layer on top, which was used as a backgate. Single-walled CNTs were prepared by the laser vaporisation method[13] at the Rice University. They were dispersed in water by sonication using sodium dodecyl sulphate (SDS) surfactant. The CNT were deposited using a combing technique, which allows us a good control of the CNT density and orientation on the substrate[14]. The silica surface was first functionalized using standard silanization technique leading to a self-assembled monolayer of aminopropyltriethoxysilane (Aldrich). The substrate was then dipped 5 minutes in the dispersion of single-walled CNTs and withdrawn at a constant velocity of 200 µm/s. The sample was thoroughly washed in distilled water in order to remove the surfactant from the nanotubes. The nanotube location was imaged by atomic force microscopy (AFM) and aligned e-beam lithography was carried out to pattern the contacts. Metal electrodes were deposited using electron-gun evaporation and a thickness of 3 nm Pd followed by 50 nm Al was used. Pd provides high-transparency contacts to the carbon nanotubes[15]. Al is a superconductor widely used in nanoscale devices, having a critical temperature of about 1.2 K. Only devices with resistance below 60 kΩ and no significant gate effect at room temperature were used for our studies. In addition to the backgate, a lateral sidegate made from a coplanar aluminium electrode was aligned to each device. All measurements presented here were performed at a cryostat temperature of about 35 mK and a magnetic field of $H_z$ = 50 mT was applied perpendicular to the electrode plane in order to suppress the superconductivity of the leads.

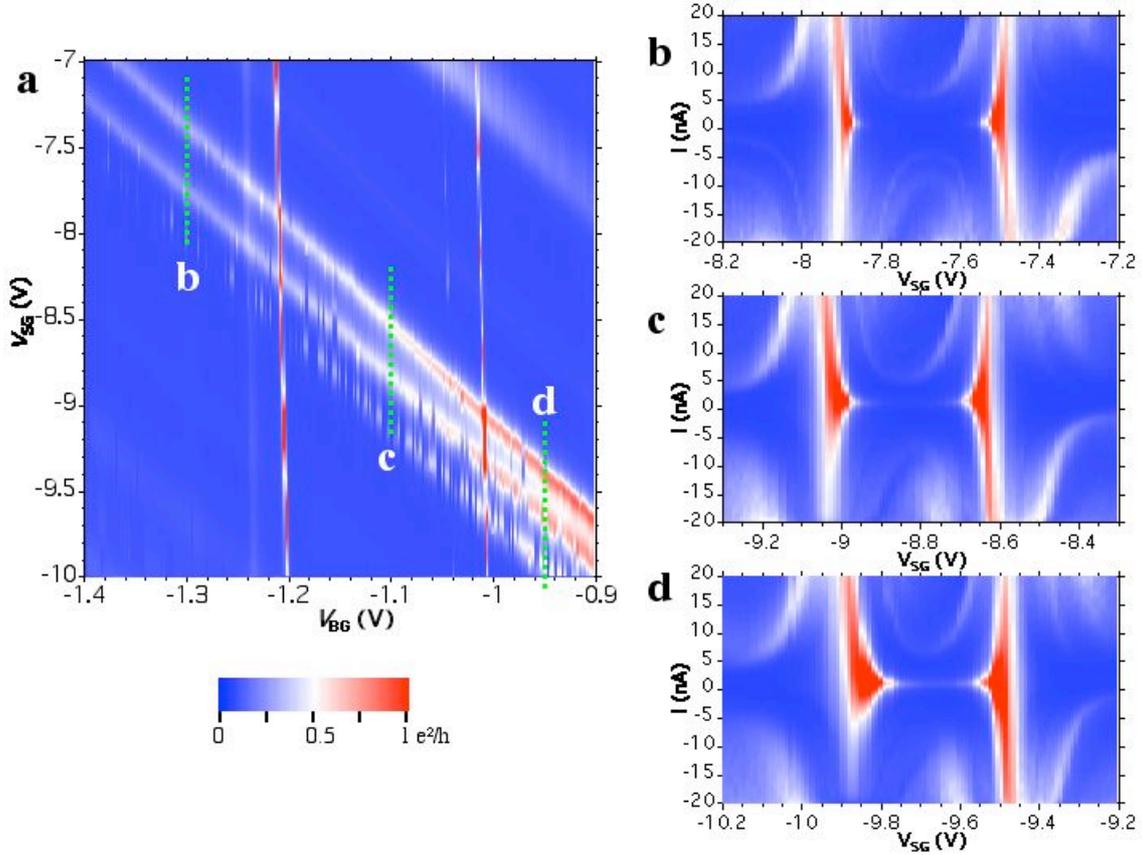

**Figure 2. Kondo effect tuning with back- and sidegates. a,** Colour-scale representation of a differential conductivity $dI/dV$ map at 34 mK as a function of the sidegate voltages $V_{SG}$ and backgate voltage $V_{BG}$. A magnetic field of $H_z$ = 50 mT was applied perpendicular to the SQUID plane in order to suppress the superconductivity of the leads. The dotted lines indicate the regions that are further studied in **b - d**. **b**, which present $dI/dV$ maps as a function of $V_{SG}$ and source-drain current $I$.

The CNT junctions can be modelled by a QD in between two metal leads. The position of the quantum levels can be tuned with a gate voltage. When a quantum level is aligned with respect to the Fermi energy of the leads, a current can flow by resonant tunnelling through the CNT. When the quantum levels are far from the Fermi energy, the current is strongly reduced. In order to characterize the electronic transport properties of our CNT-junctions, we measured the differential conductance $dI/dV$ as a function of source-drain voltage $V_{sd}$, lateral sidegate and backgate voltages $V_{SG}$ and $V_{BG}$, respectively. Whereas the source-drain voltage $V_{sd}$ shifts the Fermi energy of the left lead in respect to the right one, the gate voltages tune the position of the quantum levels with respect to both Fermi energies of the leads. In addition, the backgate electrode can vary the transparency of the QD barriers, thus permitting to change the hybridization of the QD states with the contacts.

In order to demonstrate this effect, we have measured the conductance map of $dI/dV$ versus side gate voltage $V_{SG}$ and backgate voltage $V_{BG}$ at 35mK (Fig. 2a). For a negative backgate voltage, two distinct conduction lines appear corresponding respectively to the level degeneracy of even-odd and odd-even neighbouring charge states[16-18]. When increasing $V_{BG}$, these two conduction lines shift to more negative sidegate voltages. In addition, the lines gradually approach each other and the conductance in between increases. This is better seen in Fig. 2b-d, which shows $dI/dV$ maps as a function of $V_{SG}$ and source-drain voltage $V_{ds}$ at three different backgate voltages. Fig. 2d shows a clear Kondo ridge in between the two conduction canals, whereas in Fig. 2b-c the Kondo ridge is very thin[19]. Fig. 3 shows for comparison the corresponding Kondo resonances in the middle of the Kondo ridge. These figures demonstrate that a combination of back- and sidegates allows us to tune the Kondo interaction between two levels.

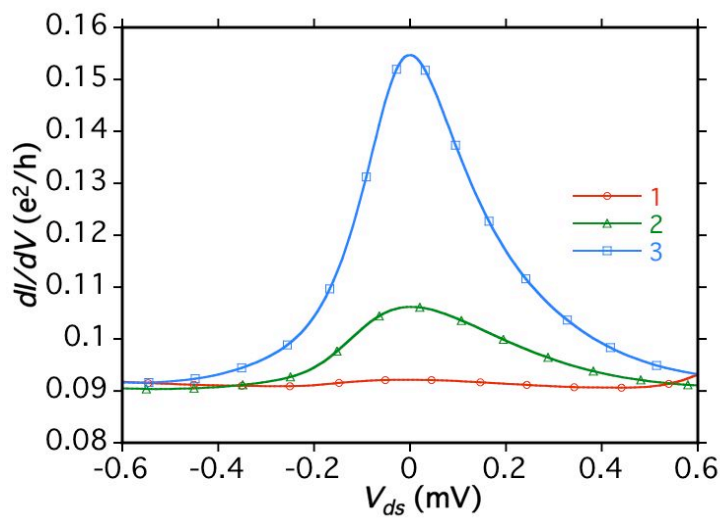

**Figure 3.** Three differential conductance versus source-drain voltage scans (1 to 3) taken in the middle of the Kondo ridge of Fig. 2b – d, respectively. It shows the gradual emergence of a Kondo resonance that is tuned with the backgate voltage.